\documentclass[10pt,journal]{IEEEtran}

\usepackage{amsthm}
\usepackage{graphicx}
\usepackage{array}
\usepackage{balance}
\usepackage{tikz}
\usepackage{pgfplots}
\usepackage{subfigure}	
\usepackage{xcolor}
\usepackage{enumerate}
\usepackage{booktabs}
\usepackage{enumerate}
\usepackage{array}
\usepackage{exscale}
\usepackage{color}
\usepackage{subfigure}
\usepackage{cite}
\usepackage{amsmath}
\usepackage{amssymb}
\usepackage{algorithm}
\usepackage{algpseudocode}
\usepackage[font=small]{caption}
\usepackage[utf8]{inputenc} 
\usepackage[T1]{fontenc} 
\usepackage{enumerate}
\usepackage{balance}

\usepackage{stfloats}

\usepackage{adjustbox}

\usepackage{mathtools}

\usetikzlibrary{arrows}

\usetikzlibrary{shapes,snakes}
\usetikzlibrary{calc}
\usetikzlibrary{patterns}
\usetikzlibrary{decorations.pathmorphing} 

\usepackage[normalem]{ulem}

\newtheorem{remark}{\textit{Remark}}

\newcommand{\TxAntenna}[3]{
\coordinate (a) at (#1,#2);
\draw[scale=(#3)] (a)--($(a)+(0.2,0)$)--($(a)+(0.2,0.7)$)--
($(a)+(0.1,0.8)$)--($(a)+(0.3,0.8)$)--($(a)+(0.2,0.7)$);
}

\newcommand{\RxAntenna}[4]{
\coordinate (a) at (#1,#2);
\draw[scale=(#3)] (a)--($(a)+(-0.2,0)$)--($(a)+(-0.2,0.7)$)--
($(a)+(-0.1,0.8)$)--($(a)+(-0.3,0.8)$)--($(a)+(-0.2,0.7)$);
}


\newcommand{\Ali}{\textcolor{black}}

\begin{document}

\title{Information Detection and Energy Harvesting Trade-off in Multi-User Secure Communication}
%
%
%

\author{
\IEEEauthorblockN{Ali Kariminezhad, Zohaib Hassan Awan, Hendrik Vogt, and Aydin Sezgin}\\
\thanks{
A. Kariminezhad, H. Vogt and A. Sezgin are with the Institute of Digital Communication Systems, Ruhr Universit\"at Bochum (RUB), Germany (emails: \{ali.kariminezhad, hendrik.vogt, aydin.sezgin\}@rub.de). Zohaib Hassan Awan is with the Institute for Theoretical Information Technology, RWTH Aachen, Germany (email: zohaib.awan@ti.rwth-aachen.de).
}}


\maketitle
\thispagestyle{empty}
\pagestyle{empty}

\vspace*{-1cm}
\begin{abstract}
The secrecy rate region of wiretap interference channels with a multi-antenna passive eavesdropper is studied under receiver energy harvesting constraints. To stay operational in the network, the legitimate receivers demand energy alongside information, which is fulfilled by power transmission and exploiting a power splitting (PS) receiver. Simultaneous wireless information and power transfer (SWIPT) results in a reduction in the secrecy rates due to the legitimate receivers PS for simultaneous information detection and energy harvesting. For this setup, lower-bounds for secure communication rate are derived without imposing any limitation at the eavesdropper processing. It turns out that in order to obtain the Pareto boundary of the secrecy rate region, smart tuning of the transmit power and receiver PS coefficient is required. Notice that, granting both, i.e., secrecy in communication as well as enabling energy harvesting, result in shrinkages in the reliable communication rate region, which is captured by this optimal tune. Hence, it is of crucial importance to investigate, which of these two services shrinks the rate region more. Interestingly, we observe that by enabling SWIPT, the convexity of the reliable rate region is preserved. However, granting secrecy for the communication results in a non-convex rate region as the interference power increases.
\end{abstract}
\section{Introduction}
Secrecy is one of the main concerns in future communication networks involving a plethora of communicating nodes. This includes wireless sensor networks (WSNs) and Internet of Things (IoTs)~\cite{Roman2005,Zou2013}. Moreover, due to a steadily increasing number of connected devices, the scarce spectrum needs to be shared among multiple communication pairs. These two factors motivate the study of the wiretap interference channel~\cite{Kalantari2015}. In this channel, multiple node pairs exchange data simultaneously over a shared spectrum, which in turn induces interference at the receivers. One way to tackle this problem is to treat the undesired signals at the legitimate receivers as noise --- popularly known as treating interference as noise (TIN). Despite, the interference is treated as noise for information detection reasons, it can be treated as a energy source for energy harvesting purposes. This way, securing the desired information assuming TIN, the legitimate users demand a particular amount of energy to stay functional during the communication process~\cite{Li2017,Chen2017}. For instance, consider a WSN with limited energy supply at the sensors. By deploying energy harvesting receivers, the energy buffer can be charged wirelessly from energy sources, e.g., solar energy, radio frequency (RF)~\cite{Habibu2014}. One aspect of this work is to investigate the concept of RF signal energy harvesting in the context of security. Now, the sensors with scarce energy supply face a trade-off in between information detection (ID) and energy harvesting (EH). Considering a single-antenna receiver, simultaneous ID and EH can be achieved by power splitting (PS). Hence, one part of received signal power undergoes the ID chain while the other part passes through the EH circuitry. Utilizing PS receivers, the required energy constraint is fulfilled by appropriate power transmission. This concept is known as simultaneous wireless information and power transmission (SWIPT)~\cite{Pan2017,Liu2013,Kariminezhad2017SPL}.

For a class of multi-user networks, the authors in~\cite{Koyluoglu2008,Liu2008} establish bounds on secure communication region. The secrecy rate region of the model that we study with EH demands is a function of --- i)  the transmit power,  and ii) the receive PS coefficients. Therefore, to establish the secrecy rate region with EH, it is pivotal to study the joint interaction between transmit power, and, receive PS coefficients at the legitimate pairs. Thus, the optimal design of these parameters captures the trade-off between secure communication rates and harvested energies.

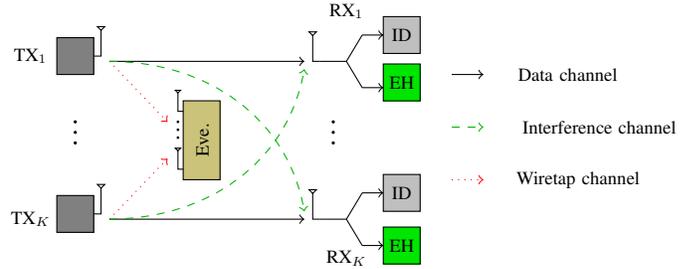
\begin{figure}{
		\tikzset{every picture/.style={scale=0.7}, every node/.style={scale=0.7}}
		\centering
		\begin{tikzpicture}
\node at (-0.5,0.35){$\text{TX}_1$};
\draw[fill=gray!100!black] (0,0) rectangle (0.7,0.7);
\TxAntenna{0.7}{0.35}{0.7};
\draw[fill=black] (0.35,-0.9) circle (0.02cm);
\draw[fill=black] (0.35,-1.1) circle (0.02cm);
\draw[fill=black] (0.35,-1.3) circle (0.02cm);

\node at (-0.5,-2.75){$\text{TX}_K$};
\draw[fill=gray!100!black] (0,-3) rectangle (0.7,-2.3);
\TxAntenna{0.7}{-2.65}{0.7};

\draw[fill=gray!50!yellow] (2.4,-2) rectangle (3.1,-0.5);
\RxAntenna{2.4}{-1.8}{0.5};
\draw[fill=black] (2.3,-1.2) circle (0.015cm);
\draw[fill=black] (2.3,-1.05) circle (0.015cm);
\draw[fill=black] (2.3,-0.9) circle (0.015cm);
\RxAntenna{2.4}{-0.7}{0.5};
\node[rotate=90] at (2.77,-1.25){ Eve.};

\RxAntenna{5}{0.25}{0.7};

\RxAntenna{5}{-2.75}{0.7};

\draw (5,0.25)--(5.5,0.25);
\draw (5.5,0.25)--(5.8,0.75);
\draw[->] (5.8,0.75)--(6.2,0.75);
\draw (5.5,0.25)--(5.8,-0.25);
\draw[->] (5.8,-0.25)--(6.2,-0.25);
\draw[fill=gray!50!white] (6.2,0.4) rectangle (6.9,1.1);
\node at (6.55,0.75){ID};
\draw[fill=green!90!black] (6.2,-0.5) rectangle (6.9,0.2);
\node at (6.55,-0.15){EH};

\node at (5.5,1.2){$\text{RX}_1$};
\node at (5.5,-3.5){$\text{RX}_K$};

\draw[fill=black] (5.25,-0.9) circle (0.02cm);
\draw[fill=black] (5.25,-1.1) circle (0.02cm);
\draw[fill=black] (5.25,-1.3) circle (0.02cm);

\draw (5,-2.75)--(5.5,-2.75);
\draw (5.5,-2.75)--(5.8,-2.25);
\draw[->] (5.8,-2.25)--(6.2,-2.25);
\draw (5.5,-2.75)--(5.8,-3.25);
\draw[->] (5.8,-3.25)--(6.2,-3.25);
\draw[fill=gray!50!white] (6.2,-2.6) rectangle (6.9,-1.9);
\node at (6.55,-2.25){ID};
\draw[fill=green!90!black] (6.2,-3.6) rectangle (6.9,-2.9);
\node at (6.55,-3.25){EH};
\draw[->,dotted,red] (1,0.25)--(2.1,-0.9);
\draw[->,black] (1,0.25)--(4.7,0.25);

\draw[->,dotted,red] (1,-2.75)--(2.1,-1.6);
\draw[->,black] (1,-2.75)--(4.7,-2.75);

\draw[->,dashed,green!70!black] (1,0.25) to[out=0,in=100] (4.7,-2.6);
\draw[->,dashed,green!70!black] (1,-2.75) to[out=0,in=-100] (4.7,0.1);

\draw[->] (7.5,0)--(8.1,0);\node at (9.7,0){Data channel};
\draw[dashed,->,green!70!black] (7.5,-1)--(8.1,-1);\node at (10.3,-1){Interference channel};
\draw[dotted,->,red] (7.5,-2)--(8.1,-2);\node at (9.9,-2){Wiretap channel};
\end{tikzpicture}
		\caption{$K$-user wiretap interference channel with power splitting structure for simultaneous information detection and energy harvesting.}
		\label{sysModel}
	}
\end{figure}
\subsection{Contribution}
In this paper, we investigate the secrecy rate region of the wiretap interference channel with EH legitimate users. Characterizing the secrecy capacity of this channel is challenging. Thus, we focus our attention to develop secure rate lower-bounds. For both secure communication and energy harvesting purposes, the transmit power, and receive PS coefficients are optimized jointly in order to capture the trade-off between secure rates and energy demands. This optimization problem is a non-convex problem. Interestingly, this problem turns out to be a signomial program (SP), which in this work is approximated by a geometric program (GP). We propose a polynomial-time algorithm to solve this problem iteratively, where the approximation gap reduces at each iteration. The optimal solution sheds light on the rate loss due to secrecy constraint, meanwhile harvesting a certain amount of RF energy.

\subsection{Notation}
Throughout the paper, we denote vectors in boldface lower-case letters while the matrices are expressed in boldface upper-case. The differential entropy of a random variable, $x$ is denoted by $\mathtt{h}(x)$ and the mutual information between two random variables, $x$ and $y$ is denoted by $\mathtt{I}(x;y)$. $\mathbf{a}\leq\mathbf{b}$ represents element-wise comparison between vectors $\mathbf{a}$ and $\mathbf{b}$. Suppose $\mathbf{x}$ is a vector of $K$ elements, i.e., $\mathbf{x}=[x_1,\cdots,x_K]^T$, then the vector $\mathbf{x}_{k+1}^{K}=[x_{k+1},\cdots,x_K]^{T}$. Moreover, $\mathbf{x}\backslash x_k$ represents vector $\mathbf{x}$ excluding the $k$--th element, i.e., $\mathbf{x}\backslash x_k=[x_1,\cdots,x_{k-1},x_{k+1},\cdots,x_K]^{T}$.

\section{System Model}\label{Sec:SystemModel}
Consider a wiretap interference channel, where the legitimate users are equipped with single antenna and the eavesdropper is equipped with multiple antennas as shown in~Fig.\ref{sysModel}. The baseband signal model at the receivers is given by 
\begin{align}
y_k&=h_{kk}x_k+\sum_{\substack{j=1\\ j\neq k}}^{K}h_{kj}x_j+w_k,\quad \forall k\in\mathcal{K},\\
\mathbf{y}_{\text{E}}&=\sum_{j=1}^{K}\mathbf{h}_{\text{E}j}x_j+\mathbf{w}_{\text{E}},\label{EveAa}
\end{align}
where the set of legitimate users is denoted by $\mathcal{K}=\{1,\cdots,K\}$. The received signal at $k$-th legitimate receiver and the eavesdropper is given by $y_k\in\mathbb{C}$ and $\mathbf{y}_{\text{E}}\in\mathbb{C}^{M}$, respectively. The transmit signal from the $k$-th legitimate transmitter is assumed to be taken from a Gaussian codebook and is denoted by $x_k\in\mathbb{C}$. We assume that, the transmit signals from the legitimate users follow independent and identical distribution. In this paper, we distinguish between the antenna noise (due to impedance mismatch) and the processing noise (due to analog/digital signal processing tasks) at the receivers. We model these noise entities as zero-mean additive white Gaussian noise (AWGN). Thus, the antenna noise at the $k$-th legitimate receiver and the eavesdropper are given by $w_k\sim\mathcal{CN}\left(0,\varrho^2_k\right)$ and $\mathbf{w}_{\text{E}}\sim\mathcal{CN}\left(\mathbf{0},\varrho^2_\text{E}\mathbf{I}_M\right)$, respectively. Channel realization from  the $j$-th transmitter to the $k$-th receiver is given by $h_{kj}\in\mathbb{C}$, and the channel from the $j$-th user to the eavesdropper is given by $\mathbf{h}_{\text{E}j}\in\mathbb{C}^{M}$. The channel state information is known globally.

%
%
The malicious eavesdropper aims at wiretapping the signals from the legitimate users. In the information decoding (ID) chain, the received signal experiences extra noise, which is originated from imperfect A/D conversion and digital processing tasks. Taking this noise term  into account, we obtain
\begin{align}
\mathbf{y}_{\text{E}}&=\sum_{j=1}^{K}\mathbf{h}_{\text{E}j}x_j+\mathbf{w}_{\text{E}}+\mathbf{n}_{\text{E}},\label{EveA}
\end{align}
where $\mathbf{n}_{\text{E}}\sim\mathcal{CN}\left(\mathbf{0},\sigma^2_\text{E}\mathbf{I}_M\right)$ represents the processing noise at the eavesdropper.
As mentioned earlier, the legitimate receivers harvest the energy from the RF signal by PS structure. Therefore, a portion of the received signal power undergoes the information detection (ID) chain and the other portion passes through the energy harvesting (EH) circuitry. The signal passed to the ID chain and EH circuitry is given by
\begin{align}
y_{\text{ID}_k}&=\sqrt{\eta_k}\bigg(h_{kk}x_k+\sum_{\substack{j=1\\ j\neq k}}^{K}h_{kj}x_j+w_k\bigg)+n_k,\quad \forall k\in\mathcal{K},\\
y_{\text{EH}_k}&=\sqrt{1-\eta_k}\bigg(\sum_{j=1}^{K}h_{kj}x_j+w_k\bigg),\quad \forall k\in\mathcal{K},
\end{align}
\Ali{where the processing noise at the $k$-th legitimate receiver is denoted by $n_k\sim\mathcal{CN}\left(0,\sigma^2_k\right)$.}
The achievable information rate and harvested energy are then given by
\begin{align}
R_k&=\mathtt{I}(x_k;y_{\text{ID}_k})\nonumber\\
&=\log\bigg(1+\frac{\eta_k p_k|h_{kk}|^{2}}{\sigma^{2}_k+\eta_k(\varrho^{2}_k+\sum_{\substack{j=1\\ j\neq k}}^{K}p_j|h_{kj}|^{2})} \bigg),\ \forall k\in\mathcal{K},\label{RateK}\\
E_k&=(1-\eta_k)\bigg(\sum_{j=1}^{K}p_j|h_{kj}|^{2}+\varrho^{2}_k\bigg),\quad \forall k\in\mathcal{K},\label{EnergyK}
\end{align}
respectively, which guarantees reliable communication with arbitrarily small decoding error. Notice that $\eta_k$ is the PS coefficient at the $k$-th legitimate receiver. The transmit power from the $k$-th transmitter is denoted by $p_k := \mathbb{E}[|x_k|^{2}]$.
\vspace{.5em}
\begin{remark}
In the interference-limited regime, i.e., for $\eta\sum_{\substack{j=1\\ j\neq k}}^{K}p_j|h_{kj}|^{2}\gg \sigma^{2}_k,\ \forall k\in\mathcal{K}$, the achievable communication rate reduction by PS is \textit{negligible}.
\end{remark}
\vspace{.5em}
\noindent For secure communication, the superposition region is given by~\cite{Wyner1975,Tekin2008}
\begin{align}
\sum_{k\in\mathcal{S}}R^{\text{s}}_k\leq\left[\sum_{k\in\mathcal{S}}
\mathtt{I}(x_k;y_{\text{ID}_k})-\mathtt{I}(\mathbf{x}_\mathcal{S};\mathbf{y}_\text{E})\right]^+,\quad\forall\mathcal{S\subseteq\mathcal{K}},
\label{SumSecRate01}
\end{align}
where the secrecy rate for the $k$-th user is denoted by $R^{\text{s}}_k$. Notice that $\mathcal{S}$ is any subset of $\mathcal{K}$, i.e.,  $\mathcal{S\subseteq\mathcal{K}}$.
\begin{remark}
With $K$ legitimate users, there exists ${K}\choose{1}$ individual secrecy rates constraints and $\sum_{k=2}^{K} {{K}\choose{k}}$ sum secrecy rates constraints.
\end{remark}
The negative term in the RHS in~\eqref{SumSecRate01} can be written as
\begin{align}
\mathtt{I}(\mathbf{x}_\mathcal{S};\mathbf{y}_\text{E})=\sum_{k\in\mathcal{S}}\mathtt{I}(x_k;\mathbf{y}_\text{E}|\mathbf{x}_{k+1}^{|\mathcal{S}|}):=\sum_{k\in\mathcal{S}}R_{\text{E}_k}.\label{MI01}
\end{align}
By definition, we have $\mathtt{I}(x_k;\mathbf{y}_\text{E}|\mathbf{x}_{k+1}^{|\mathcal{S}|})=\mathtt{I}(x_k;\mathbf{y}_\text{E})$ if $k+1>|\mathcal{S}|$. Furthermore, we define $\mathbf{x}_{k+1}^{|\mathcal{S}|}=\mathbf{x}_{k+1}$, if $k+1=|\mathcal{S}|$. With Gaussian codebooks, the term in~\eqref{MI01} is given by~\cite{Varanasi1997},
\begin{align}
&\sum_{k\in\mathcal{S}}R_{\text{E}_k}=\nonumber\\
&\log\det\left(\mathbf{I}_M+\sum_{k\in\mathcal{S}}\frac{p_k\mathbf{h}_{\text{E}_k}
\mathbf{h}^{H}_{\text{E}_k}}{\sigma^{2}_\text{E}+\varrho^{2}_\text{E}}\left(\mathbf{I}_M+\sum_{l\in\mathcal{S}^c}\frac{p_l\mathbf{h}_{\text{E}_l}
\mathbf{h}^{H}_{\text{E}_l}}{\sigma^{2}_\text{E}+\varrho^{2}_\text{E}}\right)^{-1}\right).
\end{align}
\begin{figure}
\centering
\tikzset{every picture/.style={scale=0.95}, every node/.style={scale=0.75}}%
\begin{tikzpicture}

\begin{axis}[%
xmin=0,
xmax=1,
xlabel={$R^{s}_1$ (bits/channel use)},
xmajorgrids,
xtick={0,0.2,0.4,0.6,0.8,1},
ymin=0,
ymax=1,
ylabel={$R^{s}_2$ (bits/channel use)},
ymajorgrids,
ytick={0,0.2,0.4,0.6,0.8,1},
legend style={at={(axis cs: 0,0)},anchor=south west,draw=black,fill=white, fill opacity=0.8,legend cell align=left}
]

\addplot [color=red,dashed]
  table[row sep=crcr]{1 0\\
  0.9689   0.1213\\
  0.9395 0.2224\\
  0.9116 0.3081\\
  0.8853 0.3819\\
  0.8603 0.4461\\
  0.8365 0.5025\\
  0.8139 0.5525\\
  0.7923 0.5972\\
  0.7717 0.6374\\
  0.7208 0.6688\\
  0.6645 0.7019\\
  0.602 0.737\\
  0.5322  0.7741\\
  0.4536 0.8136\\
  0.3642 0.8556\\
  0.2614  0.9005\\
  0.1417 0.9485\\
  0 1\\
};
\addlegendentry{$\left(R^s_1,R^s_2\right)=\left(R_1-\hat{R}_{\text{E}_1},R_2-\hat{R}_{\text{E}_2}\right)$};

\addplot [color=green!50!black,dashdotted]
  table[row sep=crcr]{0 1\\
  0.1213  0.9689\\
  0.2224 0.9395\\
  0.3081 0.9116\\
  0.3819 0.8853\\
  0.4461  0.8603\\
  0.5025 0.8365\\
  0.5525 0.8139\\
  0.5972 0.7923\\
  0.6374 0.7717\\
  0.6688 0.7208\\
  0.7019 0.6645\\
  0.737 0.602\\
  0.7741  0.5322\\
  0.8136 0.4536\\
  0.8556 0.3642\\
  0.9005 0.2614\\
  0.9485 0.1417\\
  1 0\\
};
\addlegendentry{$\left(R^s_1,R^s_2\right)=\left(R_1-\tilde{R}_{\text{E}_1},R_2-\tilde{R}_{\text{E}_2}\right)$};

\addplot [color=black,solid,mark=o,mark options={solid}]
  table[row sep=crcr]{1 0\\
  0.9689   0.1213\\
  0.9395 0.2224\\
  0.9116 0.3081\\
  0.8853 0.3819\\
  0.8603 0.4461\\
  0.8365 0.5025\\
  0.8139 0.5525\\
  0.7923 0.5972\\
  0.7717 0.6374\\
  0.6374 0.7717\\
  0.5972 0.7923\\
  0.5525 0.8139\\
  0.5025 0.8365\\
  0.4461  0.8603\\
  0.3819 0.8853\\
  0.3081 0.9116\\
  0.2224 0.9395\\
  0.1213  0.9689\\
  0 1\\
};
\addlegendentry{Time sharing};

\end{axis}
\end{tikzpicture}%
\caption{Achievable secrecy rate region of the symmetric two-user wiretap interference channel. Antenna and processing noise variance are set to $\sigma_k=\varrho_k=0.25,\ \forall k\in\{1,2\},\sigma_\text{E}=
\varrho_\text{E}=0.25$. Transmit power budget is $\mathbf{p}_\text{max}=\mathbf{1}$. $|h_{kk}|=1$, $\|\mathbf{h}_{\text{E}_k}\|=0.5,\forall k\in\{1,2\}$.}
\label{fig:TightLowerBoundA}
\end{figure}
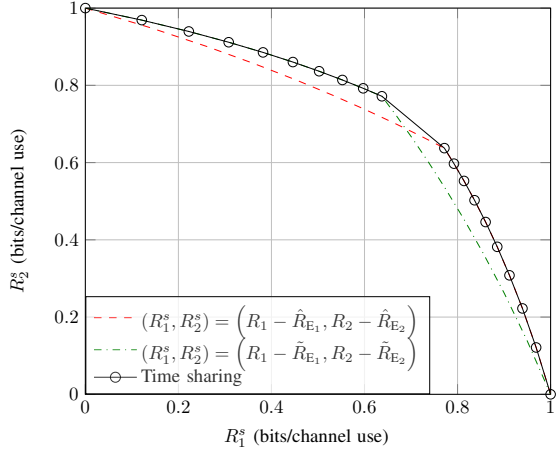 
\begin{figure*}[b]
\hrule
\begin{align}
R^s_1+R^s_2&\leq \left[\underbrace{\log\bigg(1+\frac{\eta_1 p_1|h_{11}|^{2}}{\sigma^{2}_1+\eta_1(\varrho^{2}_1+p_2|h_{12}|^{2})} \bigg)}_{R_1} + \underbrace{\log\bigg(1+\frac{\eta_2 p_2|h_{22}|^{2}}{\sigma^{2}_2+\eta_2(\varrho^{2}_2+p_1|h_{21}|^{2})} \bigg)}_{R_2}-\sum_{k=1}^{2}R_{\text{E}_k}\right]^{+},\label{eq:sumSecRateConsA}
\end{align}
where
\begin{align}
&\sum_{k=1}^{2}R_{\text{E}_k}=\log\det\left(\mathbf{I}_M+\frac{p_1\mathbf{h}_{\text{E}_1}
\mathbf{h}^{H}_{\text{E}_1}+p_2\mathbf{h}_{\text{E}_2}
\mathbf{h}^{H}_{\text{E}_2}}{\bar{\sigma}^2_\text{E}}\right)\nonumber\\
&=\log\left(1+\frac{p_1}{\bar{\sigma}^2_\text{E}}\mathbf{h}^{H}_{\text{E}_1}\mathbf{Q}^{-1}_1\mathbf{h}_{\text{E}_1}\right)+\log\left(1+\frac{p_2}{\bar{\sigma}^2_\text{E}}\|\mathbf{h}_{\text{E}_2}\|^2\right)=:\tilde{R}_{\text{E}_1}+\tilde{R}_{\text{E}_2}\nonumber\\
&=\log\left(1+\frac{p_1}{\bar{\sigma}^2_\text{E}}\|\mathbf{h}_{\text{E}_1}\|^2\right)+\log\left(1+\frac{p_2}{\bar{\sigma}^2_\text{E}}\mathbf{h}^{H}_{\text{E}_2}\mathbf{Q}^{-1}_2\mathbf{h}_{\text{E}_2}\right)=:\hat{R}_{\text{E}_1}+\hat{R}_{\text{E}_2}.\label{eq:sumSecRateConsAA}
\end{align}
\end{figure*}
By plugging~\eqref{RateK} and~\eqref{MI01} in~\eqref{SumSecRate01}, we get
\begin{align}
\sum_{k=1}^{K}R^{\text{s}}_k\leq\left[\sum_{k=1}^{K}
R_k-R_{\text{E}_k}\right]^+.\label{SumSecRate01A}
\end{align}
We define $\bar{\sigma}^2_\text{E}=\sigma^{2}_\text{E}+\varrho^{2}_\text{E}$.\\
{\textit{Example:}} Consider the case of two legitimate users, $K=2$. Then, the achievable individual secrecy rates are constrained by
\begin{align}
R^s_1\leq & \Bigg[\log\bigg(1+\frac{\eta_1 p_1|h_{11}|^{2}}{\sigma^{2}_1+\eta_1(\varrho^{2}_1+p_2|h_{12}|^{2})} \bigg) -\nonumber\\ &\log\left(1+\frac{p_1}{\bar{\sigma}^2_\text{E}}\mathbf{h}^{H}_{\text{E}_1}\mathbf{Q}^{-1}_1\mathbf{h}_{\text{E}_1}\right)\Bigg]^{+},\label{eq:IndSecRateAA}\\
R^s_2\leq &\Bigg[\log\bigg(1+\frac{\eta_2 p_2|h_{22}|^{2}}{\sigma^{2}_2+\eta_2(\varrho^{2}_2+p_1|h_{21}|^{2})} \bigg) -\nonumber\\ &\log\left(1+\frac{p_2}{\bar{\sigma}^2_\text{E}}\mathbf{h}^{H}_{\text{E}_2}\mathbf{Q}^{-1}_2\mathbf{h}_{\text{E}_2}\right)\Bigg]^{+},\label{eq:IndSecRateBA}
\end{align}
where the matrices $\mathbf{Q}_1$ and $\mathbf{Q}_2$ are given by
\begin{align}
\mathbf{Q}_1=\mathbf{I}_M+\frac{p_2\mathbf{h}_{\text{E}_2}
\mathbf{h}^{H}_{\text{E}_2}}{\bar{\sigma}^2_\text{E}},\\
\mathbf{Q}_2=\mathbf{I}_M+\frac{p_1\mathbf{h}_{\text{E}_1}
\mathbf{h}^{H}_{\text{E}_1}}{\bar{\sigma}^2_\text{E}},
\end{align}
respectively. Notice that, the matrices $\mathbf{Q}_1$ and $\mathbf{Q}_2$ coincide with the interference-plus-noise covariance matrices at the eavesdropper. Moreover, the sum secrecy rate is constrained by~\eqref{eq:sumSecRateConsA} as given at the bottom of the page.

Therefore, achievable individual secrecy rates satisfying the sum-rate constraint~\eqref{eq:sumSecRateConsA} are from the set $\mathcal{R}^s$ as
\begin{align}
\left(R^s_1,R^s_2\right)\in\mathcal{R}^s=&\Bigg\{\left([R_1-\tilde{R}_{\text{E}_1}]^{+},[R_2-\tilde{R}_{\text{E}_2}]^{+}\right),\nonumber\\
&\left([R_1-\hat{R}_{\text{E}_1}]^{+},[R_2-\hat{R}_{\text{E}_2}]^{+}\right)\Bigg\},
\end{align}
which also satisfy the constraints~\eqref{eq:IndSecRateAA},~\eqref{eq:IndSecRateBA}. Notice that the set $\mathcal{R}^s$ consists of $2$ elements for $K=2$. Hence, the convex-hull of these rate tuples in $\mathcal{R}^s$ is also achievable. Assuming zero EH constraints at the legitimate users, the achievable secrecy rate region is depicted in~Fig.~\ref{fig:TightLowerBoundA}.  This set involves $K!$ elements for any arbitrary $K$. Any element of this set yields an achievable set of secrecy rates lower-bound.

\subsection{Achievable Secrecy Rate Region}\label{Sec:AchievableSecRateRegion}
For simplicity in explanation and preserving clarity, we elaborate the secrecy rate region acquisition process for the case that $K=2$. The proposed procedure can be readily extended for arbitrarily large $K$. Recall that for $K=2$, we have $|\mathcal{R}^\mathrm{s}|=2$. Obtaining an element of the set $\mathcal{R}^\mathrm{s}$ as $(R^\mathrm{s}_1,R^\mathrm{s}_2)$, the secrecy rate region can be obtained by formulating the following weighted max-min optimization problem
\begin{subequations}\label{A1}
\begin{align}
\max_{\boldsymbol{\eta},\mathbf{p}}\quad & \min \Big(\frac{R^{\text{s}}_1}{\alpha_1},\frac{R^{\text{s}}_2}{\alpha_2}\Big)\tag{\ref{A1}}\\
\text{subject to}\quad & E_k \geq \psi_k,\quad\forall k\in\{1,2\},\label{A11}\\
& \mathbf{p}\leq\mathbf{p}_\text{max},\label{A12}\\
& \mathbf{0}\leq\boldsymbol{\eta}\leq\mathbf{1},\label{A13}
\end{align}
\end{subequations}
where $\mathbf{p}=[p_1,p_2]$ and $\boldsymbol{\eta}=[\eta_1,\eta_2]$. Moreover, $\mathbf{p}_\text{max}=[p_{1_\text{max}},p_{2_\text{max}}]$ is the power budget available at the users. The energy harvesting constraint is given by~\eqref{A11}, where the energy demand of the $k$-th user is represented by $\psi_k$. The weight vector $\boldsymbol{\alpha}=[\alpha_1,\alpha_2]$ is determined apriori with $0\leq\alpha_k\leq 1,\ \forall k$ and $\|\boldsymbol{\alpha}\|_1=1$. Notice that, with each realization of $\boldsymbol{\alpha}$ we obtain a secure achievable rate on the Pareto boundary of the secrecy rate region. Hence, solving problem~\eqref{A1} for each realization of $\boldsymbol{\alpha}\in\mathbb{R}^{2}$ with a predefined resolution, delivers the secrecy rate region~\cite{Ottersten2014}. Problem~\eqref{A1} is a non-convex problem. This is due to the non-convexity of the objective function, which is the division of non-convex functions. By defining an auxiliary variable $\beta=\min \Big(\frac{R^{\text{s}}_1}{\alpha_1},\frac{R^{\text{s}}_2}{\alpha_2}\Big)$, we transfer the objective function into the constraint set. Then we obtain,
\begin{subequations}\label{A2}
\begin{align}
\max_{\beta, \boldsymbol{\eta}, \mathbf{p}}\quad &\beta \tag{\ref{A2}}\\
\text{subject to}\quad  &\beta\alpha_k \leq R^{\text{s}}_k,\quad \forall k\in\{1,2\},\label{A21}\\
&\eqref{A11},\eqref{A12},\eqref{A13}.
\end{align}
\end{subequations}
We can reformulate constraints~\eqref{A21} as given in~\eqref{RateSignomial} and~\eqref{RateSignomialB}, where $\lambda=2^{\beta}$. Since $\lambda$ is monotonically increasing as a function of $\beta$ in optimization problem~\eqref{A2}, we replace $\beta$ with $\lambda$.
\begin{figure*}[t]
\begin{align}
\frac{\lambda^{\alpha_1}\left(\bar{\sigma}^2_\text{E}+p_1\mathbf{h}^H_{\text{E}_1}\mathbf{Q}^{-1}_1\mathbf{h}_{\text{E}_1}\right)\left(\sigma^2_1+\eta_1\varrho^2_1+\eta_1p_2|h_{12}|^2 \right)}{\bar{\sigma}^2_\text{E}\left(\sigma^2_1+\eta_1\varrho^2_1+\eta_1\sum_{j=1}^{2}p_j|h_{1j}|^2\right)}\leq 1\label{RateSignomial}\\
\frac{\lambda^{\alpha_2}\left(\bar{\sigma}^2_\text{E}+p_2\|\mathbf{h}_{\text{E}_2}\|^{2}\right)\left(\sigma^2_2+\eta_2\varrho^2_2+\eta_2p_1|h_{21}|^2 \right)}{\bar{\sigma}^2_\text{E}\left(\sigma^2_2+\eta_2\varrho^2_2+\eta_2\sum_{j=1}^{2}p_j|h_{2j}|^2\right)}\leq 1\label{RateSignomialB}
\end{align}
\hrule
\end{figure*}
The energy harvesting constraints in~\eqref{A11} are reformulated as
\begin{align}
\frac{\psi_k-\sigma^2_k+\eta_k\sum_{j=1}^{2}p_j|h_{kj}|^{2}}{\sum_{j=1}^{2}p_j|h_{kj}|^{2}}\leq 1,\quad\forall k\in\{1,2\}.\label{EnergySignomial}
\end{align}
Hence, we obtain
\begin{align}
\max_{\lambda, \boldsymbol{\eta}, \mathbf{p}}\quad  \lambda \quad\quad
\text{subject to}\quad  \eqref{RateSignomial},\eqref{RateSignomialB},\eqref{EnergySignomial},\eqref{A12},\eqref{A13}. \label{A3}
\end{align}
The constraints~\eqref{RateSignomial} and~\eqref{EnergySignomial} are divisions of posynomials, which are not necessarily convex functions. This renders the weighted max-min optimization problem into a signomial program which is a NP-hard problem~\cite{Chiang2005}. Next, we propose a polynomial-time algorithm to obtain a sub-optimal solution. We approximate the denominator of the functions in~\eqref{RateSignomial} and~\eqref{EnergySignomial} with a monomial function based on the single condensation method\cite{Chiang2005}. This approximation is based on the relation between arithmetic and geometric means~\cite{Boyd2007}. For instance, the denominator of constraint~\eqref{EnergySignomial} can be approximated as
\begin{align}
\sum_{j=1}^{2}p_j|h_{kj}|^{2}\geq\prod_{j=1}^{2}\left(\frac{p_j|h_{kj}|^{2}}{c_{kj}}\right)^{c_{kj}},
\label{Approx}
\end{align}
where $c_{kj},\ \forall j$, controls the approximation gap given by
\begin{align}
B=\sum_{j=1}^{2}p_j|h_{kj}|^{2}-\prod_{j=1}^{2}\left(\frac{p_j|h_{kj}|^{2}}{c_{kj}}\right)^{c_{kj}}.
\end{align}
Inequality~\eqref{Approx} holds with equality with the optimal value of $c_{kj},\ \forall j$, which is
\begin{align}
c^{\star}_{kj}=\frac{p_j|h_{kj}|^{2}}{\sum_{j=1}^{K}p_j|h_{kj}|^{2}}.
\end{align}
Similar approximation is applied for the denominator of~\eqref{RateSignomial}. Notice that, $c^{\star}_{kj}$ is a function of the optimization parameters $\mathbf{p}$. Here, we tighten the approximation gap by optimizing over $p_j$ and $\eta_j,\ \forall j$ and using the solutions to obtain $c^{\star}_{kj}$. Notice that this $c^{\star}_{kj}$ is optimal only for the current iteration and suboptimal for the next iteration. Furthermore, note that the matrix $\mathbf{Q}_1$ in a function of $p_2$ as  $\mathbf{Q}_1=\mathbf{I}_M+\frac{p_2\mathbf{h}_{\text{E}_2}
\mathbf{h}^{H}_{\text{E}_2}}{\bar{\sigma}^2_\text{E}}$. In the proposed iterative approach, we fix this matrix based on the solution of $p_2$ from the previous iteration, i.e., at iteration $l$ we use $\mathbf{Q}^{(l)}_1=\mathbf{I}_M+\frac{p^{(l-1)}_2\mathbf{h}_{\text{E}_2}
\mathbf{h}^{H}_{\text{E}_2}}{\bar{\sigma}^2_\text{E}}$.

\section{Numerical Results}\label{Sec:NumRes}
We consider two legitimate users that are wiretapped by an eavesdropper equipped with two antennas, i.e., $K=M=2$. We assume that the antenna noise and the processing noises at the legitimate receivers and at the eavesdropper have the variances, $\varrho^{2}_k=\sigma^{2}_{k}=0.25,\ \forall k$ and $\varrho^{2}_\text{E}=\sigma^{2}_{\text{E}}=0.25$, respectively. Moreover, the power budget at the transmitters is assumed to be 1, i.e., $\mathbf{p}_\text{max}=\mathbf{1}$. The achievable secrecy rate region Pareto-boundary of the wiretap interference channel is characterized numerically based on the proposed algorithm.
\begin{figure*}
\centering
\subfigure[Weak interference regime, $|h_{kj}|=0.5,\ \forall k\in\{1,2\}, j\neq k$]{
\tikzset{every picture/.style={scale=0.95}, every node/.style={scale=0.8}}%
\begin{tikzpicture}

\begin{axis}[%
xmin=0,
xmax=1.6,
xlabel={$R_1$ (bits/channel use)},
xmajorgrids,
xtick={0,0.2,0.4,0.6,0.8,1,1.2,1.4,1.6},
ymin=0,
ymax=1.6,
ylabel={$R_2$ (bits/channel use)},
ymajorgrids,
ytick={0,0.2,0.4,0.6,0.8,1,1.2,1.4,1.6},
legend style={at={(axis cs: 1.6,1.6)},anchor=north east,draw=black,fill=white, fill opacity=0.8,legend cell align=left}
]
\addplot [color=green!40!black,solid,mark=*,mark options={solid}]
  table[row sep=crcr]{1.4126153563554e-06	1.58496201948188\\
  0.0823205218443338	1.56408992554952\\
  0.171211221942286	1.54090080925488\\
  0.26736528123968	1.51506983582106\\
  0.371556265491209	1.48622504449094\\
  0.484647916146558	1.45394370917348\\
  0.607606812637091	1.41774922780937\\
  0.741520471093087	1.37710944100909\\
  0.887624862257573	1.33143728521339\\
  1.04734938789605	1.28009369760671\\
  1.22239241993211	1.22239241993211\\
  1.28009369695044	1.04734938924907\\
  1.33143728573012	0.887624861990992\\
  1.37710943997653	0.74152047191924\\
  1.41774922787568	0.607606812603918\\
  1.45394371794899	0.48464790699806\\
  1.48622504742746	0.371556262252631\\
  1.51506984570426	0.26736526885766\\
  1.5409008210719	0.171211205552617\\
  1.56408992552826	0.082320521926658\\
  1.58496204019499	1.35200918682798e-06\\
  };
\addlegendentry{Reliable, $\mathcal
E=\{0,0\}$};

\addplot [color=blue,solid,mark=asterisk,mark options={solid}]
  table[row sep=crcr]{4.0781395777012e-06	1.1341413130581\\
  0.0595664587898648	1.1317622162536\\
  0.12541741373404	1.12875637867747\\
  0.198517396772161	1.12493175582607\\
  0.2800094280541	1.1200374891748\\
  0.371248690180841	1.11374588353865\\
  0.473839880170918	1.10562635872666\\
  0.589673832969413	1.09510854408693\\
  0.720955287484186	1.08143292370103\\
  0.870210735408415	1.06359089714437\\
  1.04026386961003	1.04026386961001\\
  1.06359089711795	0.870210735456962\\
  1.08143292348461	0.720955288016354\\
  1.09510854402222	0.589673833161365\\
  1.10562636154404	0.473839869924784\\
  1.11374589516963	0.371248639857641\\
  1.12003749901631	0.280009378202923\\
  1.12493175996643	0.198517372299984\\
  1.12875638313382	0.125417383143194\\
  1.13176221624646	0.0595664590069441\\
  1.13414131267544	4.09866533856995e-06\\
  };

\addlegendentry{Reliable, $\mathcal
  E=\{0.8,0.8\}$};

\addplot [color=black,solid,mark=o,mark options={solid}]
  table[row sep=crcr]{0	1\\
  0.106608854642153	0.959479688636523\\
  0.227263430428912	0.909053720274238\\
  0.362559300227099	0.845971697928448\\
  0.511510785955879	0.767266178590319\\
  0.767266178588783	0.511510785958606\\
  0.845971697935813	0.362559300234502\\
  0.909053720274933	0.22726343042806\\
  0.959479688636971	0.106608854641494\\
  0.959479688636971	0.106608854641494\\
  1 0\\
  };
\addlegendentry{Secure, $\mathcal
E=\{0,0\}$, optimization};

\addplot [color=red,solid]
  table[row sep=crcr]{1.53676860126239e-09	0.618235867717885\\
    0.0679382984710492	0.611444676720813\\
    0.150308400216383	0.601233594792858\\
    0.250847585615054	0.585311031649188\\
    0.373130329292121	0.559695492656704\\
  0.559695492662502	0.37313032926656\\
  0.585311031648134	0.250847585615291\\
  0.601233594794108	0.150308400207432\\
  0.611444676725041	0.0679382984660567\\
  0.618235867714526	1.56969470754831e-09\\
  };
\addlegendentry{Secure, $\mathcal
E=\{0.8,0.8\}$, optimization};

\end{axis}
\end{tikzpicture}%
\label{fig:ReliableSecureA}
}\quad\quad\quad\quad
\subfigure[Strong interference regime, $|h_{kj}|=1,\ \forall k\in\{1,2\}, j\neq k$]{
\tikzset{every picture/.style={scale=0.95}, every node/.style={scale=0.8}}%
\begin{tikzpicture}

\begin{axis}[%
xmin=0,
xmax=1.6,
xlabel={$R_1$ (bits/channel use)},
xmajorgrids,
xtick={0,0.2,0.4,0.6,0.8,1,1.2,1.4,1.6},
ymin=0,
ymax=1.6,
ylabel={$R_2$ (bits/channel use)},
ymajorgrids,
ytick={0,0.2,0.4,0.6,0.8,1,1.2,1.4,1.6},
legend style={at={(axis cs: 1.6,1.6)},anchor=north east,draw=black,fill=white, fill opacity=0.95,legend cell align=left}
]
\addplot [color=green!40!black,solid,mark=*,mark options={solid}]
  table[row sep=crcr]{2.01858757749029e-05	1.5849082514249\\
  0.0760005770492131	1.44401095491215\\
  0.147615036829616	1.32853529515463\\
  0.21699246222825	1.22962395120892\\
  0.285562642501061	1.14225056883712\\
  0.35442405667	1.06327216938272\\
  0.42452985899014	0.990569687731084\\
  0.49679825842569	0.922625342983534\\
  0.57219516668612	0.85829274876941\\
  0.651813309359666	0.796660711247636\\
  0.736965594011566	0.736965594011567\\
  0.796660711200495	0.651813309346242\\
  0.858292749473213	0.572195166844125\\
  0.922625346010612	0.49679825882572\\
  0.990569687560538	0.424529858989556\\
  1.06327216862456	0.354424056696872\\
  1.14225056955531	0.285562642444004\\
  1.22962395084902	0.216992462265002\\
  1.32853530974793	0.147615034502239\\
  1.44401095489721	0.0760005770624955\\
  1.58490825206088	2.01858443734659e-05\\
  };
\addlegendentry{Reliable, $\mathcal
E=\{0,0\}$};

\addplot [color=blue,solid,mark=asterisk,mark options={solid}]
  table[row sep=crcr]{ 0	0.922863446789654\\
  0.0674455950199555	0.922863446789654\\
  0.162466817329093	0.920645280881376\\
  0.226343621071714	0.905374480086192\\
  0.293307075640863	0.879921223276816\\
  0.363279194225972	0.847651451116123\\
  0.436558547543637	0.810751587326395\\
  0.513733888689125	0.770600832530219\\
  0.595666177265281	0.728036435144139\\
  0.683526334023581	0.683526334023554\\
  0.72803643517594	0.595666177270192\\
  0.770600832535147	0.513733888689348\\
  0.810751587325234	0.436558547543845\\
  0.847651451454296	0.363279194143154\\
  0.879921221513605	0.293307076515452\\
  0.905374480374534	0.226343620957256\\
  0.920645279955747	0.162466819550918\\
  0.922863446788689	0.0674455947844663\\
  0.922863446788689	0\\
  };
\addlegendentry{Reliable, $\mathcal
E=\{1,1\}$};

\addplot [color=black,solid,mark=o,mark options={solid}]
  table[row sep=crcr]{3.73701476674704e-06	0.999977063835348\\
  0.0741017512560123	0.666898335742331\\
  0.129773240265166	0.519083542720247\\
  0.181528511437516	0.423561601703055\\
  0.233489125526238	0.350231697597165\\
  0.288607172518694	0.288607172451801\\
  0.350232087449639	0.233489309968792\\
  0.423563022778258	0.181528802993471\\
  0.519087641997721	0.129773494621316\\
  0.66689611995628	0.0741016651453204\\
  0.9999730238545	3.54237512656397e-06\\
  };
\addlegendentry{Secure, $\mathcal
E=\{0,0\}$, optimization};

\addplot [color=red,solid]
  table[row sep=crcr]{0.00285897548082059	0.420459043252622\\
    0.0389116220792438	0.350204592466408\\
  0.102929902934684	0.240169773831441\\
  0.130608444182121	0.195910995052433\\
  0.159546963284214	0.159546963293873\\
  0.191526692174424	0.127685544730963\\
  0.229146105572633	0.0982070282884864\\
  0.277493758728557	0.0693766765070676\\
  0.350204592542193	0.0389116220516735\\
  0.420459043269837	0.00285897552667735\\
  };
\addlegendentry{Secure, $\mathcal
E=\{1,1\}$, optimization};
\end{axis}
\end{tikzpicture}%
\label{fig:ReliableSecureB}
}
\caption{Comparison between the achievable reliable and secure communication rates. Antenna and processing noise variance are set to $\sigma_k=\varrho_k=0.25,\ \forall k\in\{1,2\},\sigma_\text{E}=
\varrho_\text{E}=0.25$. Transmit power budget is $\mathbf{p}_\text{max}=\mathbf{1}$. $|h_{kk}|=1$, $\|\mathbf{h}_{\text{E}_k}\|=0.5,\forall k\in\{1,2\}$.}
\label{fig:ReliableSecureRateRegion}
\end{figure*}
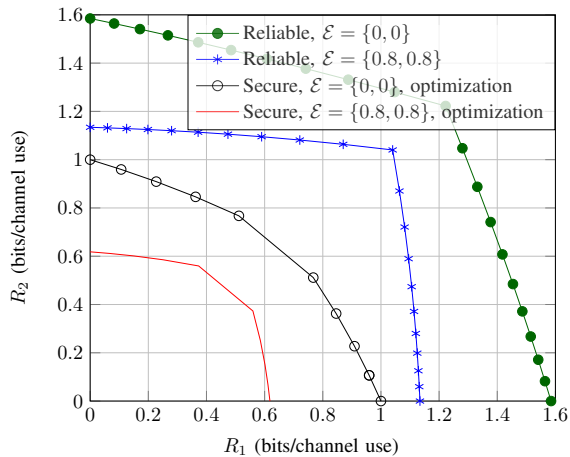
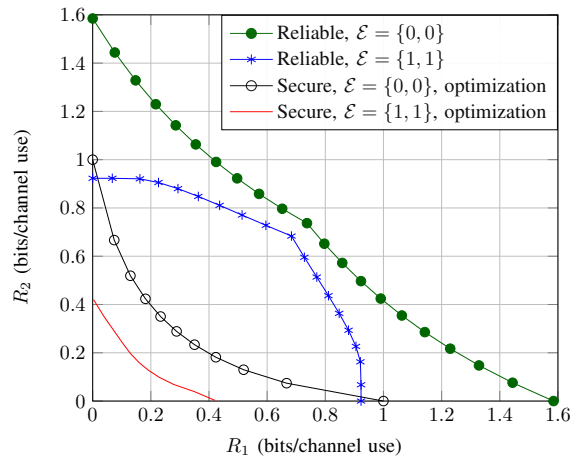 
In~Fig.~\ref{fig:ReliableSecureRateRegion}, the Pareto-boundary of the reliable and the secure communication rates are depicted. Interestingly, for weak interference channels~Fig.~\ref{fig:ReliableSecureA}, both reliable and secure communication regions are convex by optimal power allocation. Hence, there is no need for time sharing between the single-user decodable operating points. Similar observations are made, with the energy harvesting demands at the legitimate users. In the strong interference regime, power allocation does not always provide an optimal solution. According to~Fig.~\ref{fig:ReliableSecureB}, power allocation is the optimal strategy for the reliable communication with EH demands. It turns out that, time-sharing achieves a larger secrecy rate region for both cases, i.e., with and without EH demands. Evidently, stronger interference results in a lower achievable secrecy rates. Interestingly in weak interference scenarios high EH demands are not feasible. This can be observed in~Fig.~\ref{fig:ReliableSecureA}, where unity EH demands are not feasible, however these demands are feasible for stronger interference scenarios as can be seen in~Fig.~\ref{fig:ReliableSecureB}.
\section{Conclusion}\label{Sec:Conclusion}
We investigated the secrecy rate region of the wiretap interference channel with energy harvesting constraints at the legitimate users. The achievable secrecy rate region is a function of power allocation and receiver PS coefficients. Hence, optimal resource allocation is required to characterize the secrecy rate region. We formulated the secrecy rate region Pareto-boundary characterization problem as a weighted max-min optimization problem. This turns out to be a non-convex problem, for which we proposed a polynomial-time algorithm to obtain a reliable sub-optimal solution. Interestingly, we observe the shrinkage of reliable communication rate region while granting both secrecy and energy. With energy demands power transmission is required, which results in a shrinkage in the rate region, however, it stays convex. But providing secrecy, the convexity of the rate region is lost and it is optimal to perform time-sharing.

\bibliographystyle{IEEEtran}
\bibliography{reference}
\balance
\end{document}